| | |
|---|---|
| **Date:** | Jun 16, 2021 |
| **To:** | "Dimitris Assanis" dimitris.assanis@stonybrook.edu |
| **From:** | "International Communications On Heat & Mass Transfer" ichmtjm@elsevier.com |
| **Subject:** | Your Submission to the ICHMT |

Ref.: Ms. No. ICHMT-D-21-00802R1
Wood Stove Combustion Modeling and Simulation: Technical Review and Recommendations
International Communications in Heat and Mass Transfer

Dear Prof. Assanis,

I am pleased to tell you that your work has now been accepted for publication in International Communications in Heat and Mass Transfer.

It was accepted on Jun 16, 2021

We appreciate and value your contribution to International Communications in Heat and Mass Transfer. We regularly invite authors of recently published articles to participate in the peer review process. You are now part of the International Communications in Heat and Mass Transfer reviewer pool. We look forward to your continued participation in our journal, and we hope you will consider us again for future submissions

With kind regards

William Worek
Editor-in-Chief
International Communications in Heat and Mass Transfer

---

In compliance with data protection regulations, you may request that we remove your personal registration details at any time.  (Use the following URL: https://www.editorialmanager.com/ichmt/login.asp?a=r). Please contact the publication office if you have any questions.

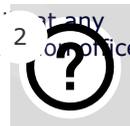

# Wood Stove Combustion Modeling and Simulation: Technical Review and Recommendations

Mahmoud Koraïem[a], Dimitris Assanis[a,b,c,*]

[a]*Mechanical Engineering, Stony Brook University, New York*
[b]*Institute for Advanced Computational Sciences, Stony Brook University, New York*
[c]*Advanced Energy Research and Technology Center, Stony Brook, New York*

ARTICLE INFO

ABSTRACT

With the rise in wood stove use in recent years, a number of experimental research efforts have been undertaken with the objective of developing cleaner, and more efficient stove designs. However, numerical modeling of stoves is still in its nascent years, thus making it imperative to start adapting more mature computational techniques, established in other combustion-based applications (e.g., combustion engines), to wood stoves. This study focuses on a critical review of the state-of-the-art in wood combustion modeling techniques and on providing recommendations for needed enhancements. In order to develop an optimum conceptual numerical model for wood stoves, the basic processes inside the stove are broken down and isolated into generic computational domains: surface reactions, gas reactions, and the fluid domain. Then state-of-the-art numerical efforts for wood stove and similar applications are broken down in a similar manner to highlight points of strength and possible improvements. Finally, based on the comparison between the different models, an improved modeling approach is proposed as a road map for future research efforts to achieve a higher fidelity wood stove numerical model.

## 1. Introduction

Wood stoves continue to increase in popularity, especially in Northeast states, with the latest report from United States (US) Energy Information Administration (EIA) showing an increase ranging from 50% to 150% over the years of 2005 to 2012 for wood heating use [1]. In addition, the EIA's 2015 Residential Energy Consumption Survey (RECS) shows that 10.5% of American households (12.5 of 118.2 million total) use wood for heating. Of those 12.5 million households, 27.2% use wood as a primary source of heat, while the remainder of homes use wood as a secondary source of heat. From an energy consumption standpoint, a total of 512.7 trillion BTUs is consumed; 60% and 40% owed to primary and secondary sources of heat, respectively [2]. Furthermore, when evaluating wood use by household income levels, wood consumption rates are fairly constant above the US median income of $50,221 [3]. However, consumption rates increase significantly and are inversely proportional with income below the US median, with the lowest income bracket of <$20K consuming the most wood.

Even though the use of wood stoves for heating is often seen as rudimentary, it is evident from the data above their use is prominent, especially as a seconding heating method. It has many perks that compete with more modern, efficient heating methods and some reasons for its popularity include:

- Wood logs, properly referred to as cord wood, can be chopped and transported in person (virtually free) or cheaply bought and easily transported with regular vehicles. It does not require infrastructure, pipelines, cables or special transport vehicles.

- Wood fuels such as cord wood are easier and safer to store on premises at larger quantities compared to liquid or gaseous fuels. They do not require any specialized storage equipment such as pressurized tanks and can be stored as simple stacked cords of wood.

- Wood is very suitable for remote areas, and as an emergency heat source in cases of natural catastrophes.

- Wood stoves are small in size and require minimal modification to households to install.

However, wood stoves and their fuels have a few major disadvantages:

- Wood has a much lower energy density than liquid and compressed gaseous fuels: ranging from a lower end of 3.5 MJ/L for Pine, Spruce and Fir wood up to 6.5 MJ/L for Hickory wood when arranged in a standard 128 cubic ft wood cord. This is compared to an energy content of 32-34 MJ/L for gasoline, 34-36 MJ/L for Diesel, and 35 MJ/L for liquefied natural gas [4, 5].

- Wood stoves have pollutants that are orders of magnitude larger than other fuel types causing an estimate of 4 million premature deaths yearly [6].

- Lastly, pellet fuels must be stored properly, with ventilation, otherwise carbon monoxide off-gassing can pose a serious health hazard [7].

The energy density consideration can be slightly improved by stacking wood in a more space efficient method, using higher energy density wood products such as dried, seasoned wood, or using wood pellets. However, the wood stove pollution consideration is the larger and more pressing issue; in addition to the carbon dioxide ($CO_2$) emissions from the combustion reaction, comparing oak combustion to propane combustion for cooking, oak emits 30-130x more particulate matter of 2.5 micron and smaller (PM2.5) and 3-30x more carbon monoxide (CO) per unit of thermal energy, in addition to black carbon and unburned hydrocarbons (UHC), and sometimes methane ($CH_4$). Nitrogen oxides ($NO_x$) emitted from wood stoves have been tested to be comparable to that of propane combustion for cooking purposes [6].

*Corresponding Author
dimitris.assanis@stonybrook.edu (D. Assanis)

Due to the extremely high concentrations of pollutants, regulations for wood stove combustion were created by the Environmental Protection Agency (EPA) to regulate wood stove emissions [8]. In consequence, research efforts focus on two main paths; reducing $CO_2$ emissions by improving stove energy overall efficiency [9], and the use of sustainable wood for combustion purposes, to reduce the overall carbon footprint during the fuel life cycle [10]. The other is improving the wood combustion efficiency and reducing emitted pollutants that are unique to wood combustion, such as excessive amounts of PM, unburned hydrocarbons, and black carbon [11].

## 2. Wood Stove Modeling

### 2.1. Importance of Numerical Modeling

Experimental investigations of wood stoves typically report measurements at the stove's boundaries (air flow in/out, emissions at the stack, etc.), and some lower resolution spatially resolved data (temperature grid using thermocouples) [10]. The data acquired from such experiments are very useful to characterize different scenarios using different fuels and operating under different conditions. However, it's not sufficient in some other scenarios, as the data/measurements are typically sparse

For example, Marabini et al. measured emissions of ultrafine particulate matter (UFP) and CO of Fir wood, Beech firewood, and Beech pellets [12]. It was found that Beech firewood had less CO emissions than Fir wood (2976 mg/MJ vs 4899 mg/MJ), while Beech pellets had more CO emissions than Fir wood (922 mg/MJ vs 216 mg/MJ). Interestingly, opposite results were found for UFP, as Beech firewood had more UFP emissions than Fir wood (67.4 mg/MJ vs 36.2 mg/MJ), while Beech pellets had less CO emissions than Fir wood (25.2 mg/MJ vs 29.7 mg/MJ).

While the previous results showed the effect of wood species and size on emissions, more data is required to actually correlate between changes in wood geometry and species, and the emissions at the stack. Pure experimental research for wood stoves cannot provide that clear correlation between results and initial or boundary conditions, and output at the stack, nor any inflection points in acquired results, without the use of multiple design points for the test, which can be - in most cases, impractical. The same applies for design optimization studies that require a multitude of different design points. Moreover, spatial data are sometimes crucial to establish the aforementioned correlation and are even harder to measure without using extremely complex test rigs. These challenges create research gaps making numerical modeling crucial for (1) combustion chamber designs investigations, (2) stove operation, control, and optimization, (3) and advanced combustion performance and emissions analysis.

### 2.2. Wood Stove Basic Model

To properly model a wood stove, one must break down the stove system into simpler components. Boundaries and processes, interacting inside a wood stove should be identified and included in the numerical model. In a wood stove, there are two main boundaries and two main zones. The fresh air inflow can usually be modeled as a boundary with a known velocity or mass flux (primary and secondary air – if it exists, are modeled similarly). The exhaust outflow can also be modeled as a known pressure boundary layer. The surface reaction zone represents the region surrounding the wood surface, and a gas reaction zone represents the combustion reactions, as shown in Figure 1.

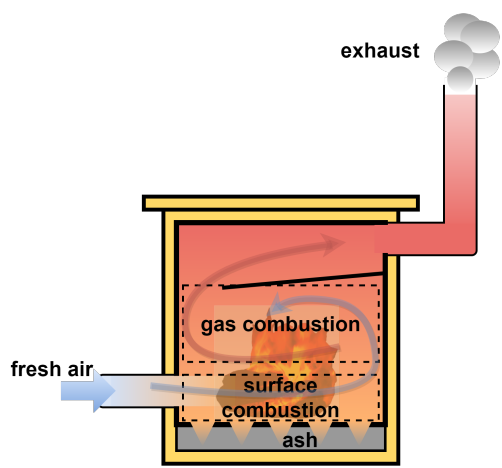

**Figure 1:** Schematic diagram of wood stove.

Reaction zone activities are dependent on local chemical species flow, concentration, and local temperature. Wood combustion main reactions are: drying, devolatilization, carbon char gasification, and combustion of volatile gases, tar and gasified char. Drying is a surface reaction where water ($H_2O$) evaporates from wood at temperatures exceeding 100°C. Devolatilization (pyrolysis) is another wood surface reaction that occurs in local oxygen ($O_2$) deficiency (equivalence ratio ($\phi$) > 1) at temperatures above 300°C, where volatile gases; CO, $CH_4$, and other hydrocarbons are emitted from the wood surface. Char left on the wood surface after drying and devolatilization is gasified; reacting with $O_2$ and combustion products ($H_2O$ and $CO_2$), forming CO and hydrogen ($H_2$) that combust in the flue gas region. All of the volatile gases and gasified char ignite at temperatures exceeding 800°C with excess $O_2$ ($\phi < 1$). Figure 2 illustrates a flow chart of surface and gas reactions in the wood stove [11].

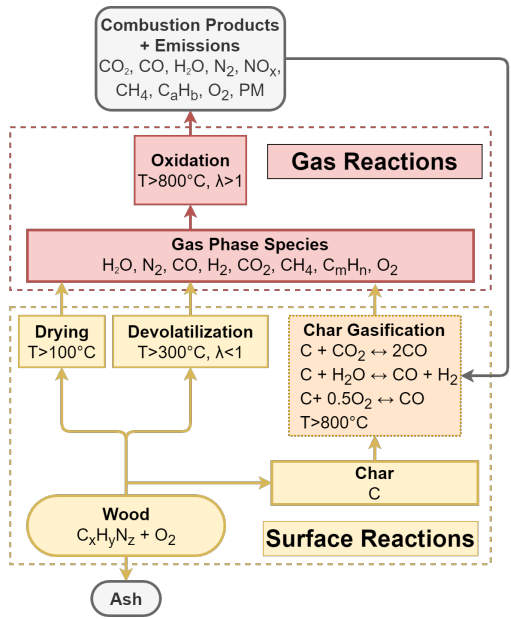

**Figure 2:** Diagram of wood surface and gas reactions [11].

To create a stove numerical model with acceptable accuracy, a high resolution 3D Computational Fluid Dynamics (CFD) model with an appropriate turbulence model is needed to properly calculate flow field spatial variables (velocity, pressure, temperature, species, etc.). Then the data should be coupled with a chemical reaction model to compute chemical species concentration and to a radiation model (to complement heat transfer calculation). From the stove and stove reaction diagrams in Figure 1 and 2, we can create a schematic diagram for a generic model to resolve a wood combustion domain, as seen in Figure 3. The model will require a two-way coupling between the flow field, and both the surface and gas reaction mechanisms.

A combustion reaction is not just a simple chemical reaction in near equilibrium conditions, but rather a complex reaction network with many intermediate chemical species and radicals reacting at varying rates in a non-equilibrium thermodynamic setting [13]. In such a setting, reaction rates vary with local temperatures (Arrhenius) and rates of change of local concentrations. The effect of flow turbulence on rate of change of species increases as the ratio of their time scales decreases, making knowledge of turbulence in a combustion zone crucial to accurately determine reaction rates [14].

**2.3. Turbulence Modeling (RANS vs. LES)**

As discussed earlier, turbulence and temperature are required to properly determine reaction rates for combustion. However there has to be a compromise between solution fidelity, computational power, and time requirement. Direct numerical simulation (DNS) can be used to fully resolve a flow field by calculating the Navier-Stokes equation for the whole domain fluid structures on all relevant length scales. However, it is so computationally expensive that it is not practical for most engineering applications. Therefore, turbulence should be modeled using one of the following two options: Reynolds Averaged Navier-Stokes equations (RANS) or Large Eddy Simulation (LES).

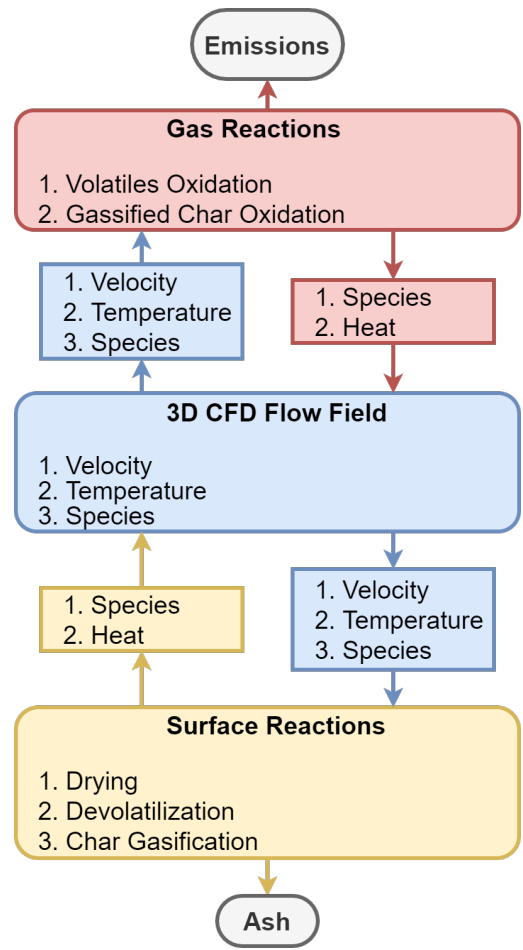

**Figure 3:** Schematic diagram of a generic wood stove numerical model

RANS models the effects of turbulence on the flow field through the addition of time-averaged variables (e.g., turbulent viscosity and dissipation in the k-$\varepsilon$ turbulence model) to the Navier-Stokes (N-S) equations. The resulting flow field calculates the time-averaged variables in the N-S equations with less computational expense, but loses instantaneous flow field values for small length/time scales [15, 16].

LES, on the other hand, works by using a spatial filter; eddies with a length scale larger than a set sub-grid size are included in the N-S equation, while eddies with smaller length scales are modeled for turbulence effect. The resulting flow field carries more flow details over the solution time domain at the expense of a much higher computational power requirement. However, due to the fluctuating nature of an LES-resolved flow field, calculating bulk domain variables (e.g., rate of heat transfer, convective boundary layers, etc.) in the solution time domain should be large enough to include multiple cycles of the observed phenomena to eliminate cyclic variability errors.

Live-images of flame instances in a wood stove have been digitally-processed to qualitatively illustrate the resolution and fidelity that can be achieved using RANS and LES modeling in contrast with the actual phenomena, at different time and length scales, as shown in Figure 4. The RANS-style processed images were created by time averaging the real flame images at three different time scales ($\delta t$, $2\delta t$, and $3\delta t$), which were varied by changing the number of frames considered. The LES-style processed images were created by applying a variable size spatial blur on pixels with a high red channel intensity (flame pixels). The spatial blur filter averaged small length scale eddies, while leaving the bulk fluid structures intact. The difference in eddy structures between the digitally-processed RANS and LES, and real flames are comparable to contours created by Som et al. when comparing RANS and LES modeled fuel particle motion from in-cylinder spray to real images [16].

### 2.4. Gas and Surface Reaction Modeling

Wood stove boundaries and chemical reactions can be modeled with varying degrees of accuracy, based on the required analysis type, available computational power, and experimental data available to validate the modeling results. For gas reactions, the simplest zero-dimensional (0D) model is based on the Arrhenius equation and is temperature dependent [15]. The temperature/time profile for the Arrhenius equation can be acquired from experimental data or from a CFD model.

Another way to represent gas phase combustion is using the Eddy Dissipation Model (EDM), which takes into account the effects of turbulence on reaction rates. EDM considers that turbulent mixing timescales are relatively larger than chemical timescales and thus the mixing of fuel, oxidizer and reaction products in the eddies will be the rate-limiting process that dominate reaction rates [17]. However, EDM neglects intermediate kinetics, inclusive of intermediate species and reactions, and thus does not capture reaction time scales. Instead EDM models chemical reactions as instantaneous phenomena and neglects the effect of turbulence on multi-step reactions [18].

Lastly, the Eddy Dissipation Concept (EDC) combines turbulent mixing from EDM and applies it to a multi-step reaction using chemical kinetics[19]. EDC can be coupled with a high resolution CFD model with detailed turbulence and with detailed chemical kinetic models to successfully resolve combustion spatially. Additionally, the combustion product species resulting from this models can be used as inputs to model emissions [15].

On the other hand, surface reactions can be modeled as simple boundaries with a mass flux of chemical species (char and devolatilization gases). The flux can be determined using experimentally measured chemical species or empirical (Arrhenius) models that calculate reaction rates using temperatures derived from (1) the CFD domain or (2) measured experimentally [20]. Alternatively, wood can be represented as a deforming boundary over the solution time domain. The surface reactions on this boundary can be modeled by releasing reactants from the boundary to interact with the CFD flow domain. An alternative, non-CFD approach to modeling cord wood pyrolysis, is to use multiple control volumes (CVs). Discretizing a wood log into cylindrical CVs allows for reaction rates to be locally resolved such that drying and shrinkage of the log during pyrolysis is accurately capture [21]. For wood chips and wood pellets, the solid fuel can be modeled as small particles with different geometric proportions [22], or as zero dimensional point particles using the discrete element method (DEM), where the domain is a very large number of particles affected by gravity, friction, and particle-to-particle impact, which is then coupled to a CFD grid [23].

## 3. Current Modeling and Simulation Approaches Applied to Improve Wood Combustion

Wood combustion models vary in their fidelity, as well as their computational expense, depending on the approach used to describe combustion and flow phenomena. In this section, examples of different approaches for modeling combustion and flow in wood stoves and other wood combustion applications, such as wood-fired boilers, furnaces, and wood chip Circular Fluidized Beds (CFB) are presented, as the modeling principles are similar.

### 3.1. Surface Flux → 3D RANS CFD Flow ↔ EDC Combustion Model

In this approach, shown in the schematic of Figure 5, surface reactions are not modeled at the wood surface; the latter is replaced with a boundary that includes surface flux of species originating from drying, devolatilization, and char gasification products. Flux species are fed to a RANS CFD model which is coupled with EDC to model combustion spatially.

Examples detailing this approach include the work of Motyl et al. that investigated new designs for wood stoves [10]. They replaced surface reaction model with a species flux from wood boundary that includes a simplified devolatilization product and char particles. A CFD model with RANS based k-$\varepsilon$ turbulence was used. A simplified seven (7) species, six (6) reactions chemical kinetic mechanism was employed.

In addition, Kalla et al. used similar approach for modeling wood log stoves[24]. They replaced surface reactions with a flux of reactants. Char gasification was modeled as a gas reaction, rather than a surface one, by introducing wood char particles in the domain from the wood boundary along with drying and pyrolysis products. A CFD model with a RANS k-$\varepsilon$ turbulence model was used. Combustion was assumed to be homogeneous along the log. A simplified eight (8) species, six (6) reactions chemical kinetic mechanism was resolved to resolve gas

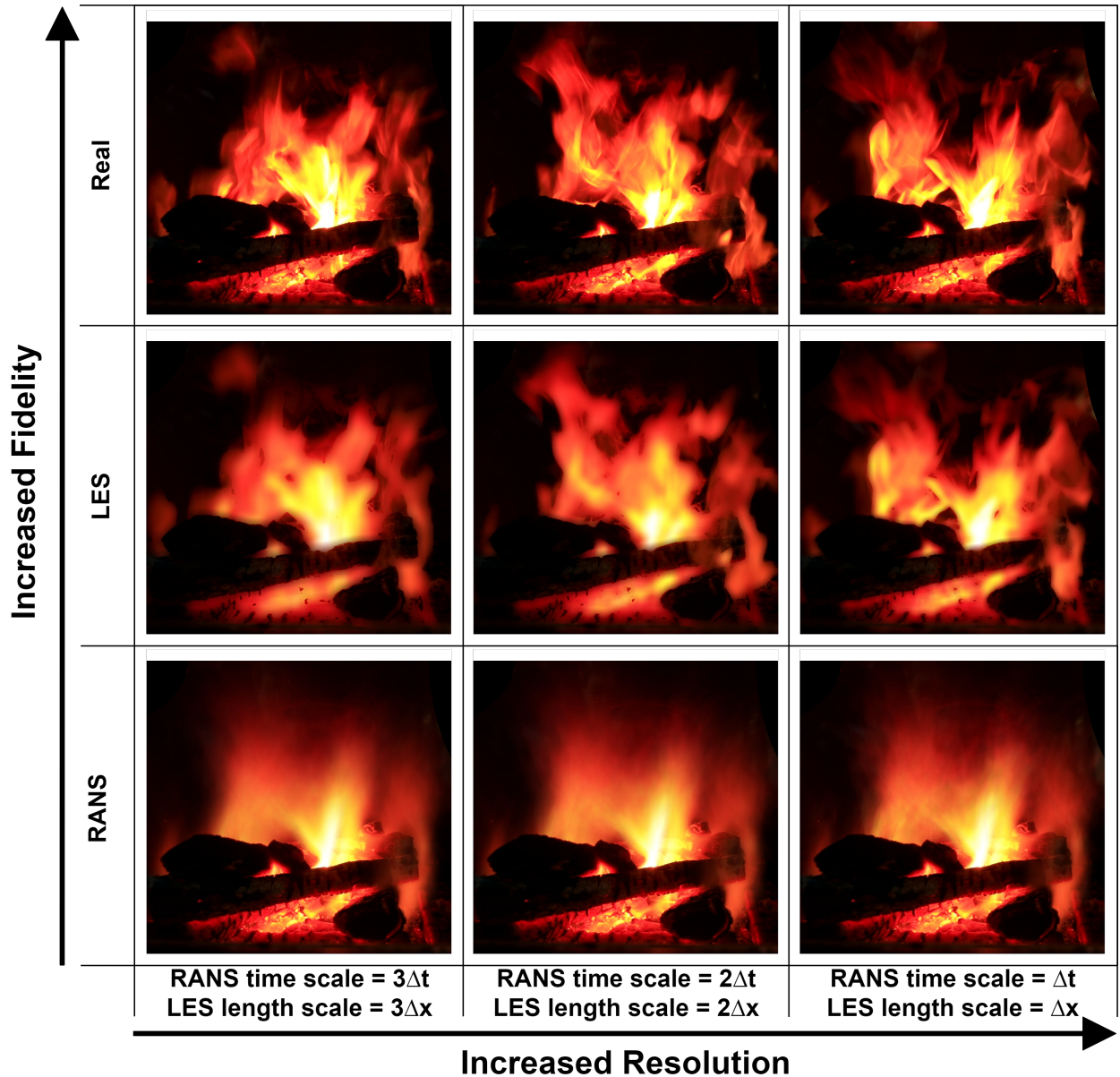

**Figure 4:** Live-images of flame instances in a wood stove compared to digitally-processed features that can be captured using RANS and LES modeling.

phase reactions.

Following the same modeling technique, Nesiadis et al. modeled a wood log boiler [25]. The created the model with five (5) species, a two (2) reaction chemical kinetic mechanism, and a domain of three zones: (1) ash formation, (2) volatile release, and (3) CFD flow field. They opted to use char gasification species instead of char particles in the gas reactions.

A final example of the surface flux/RANS/EDC approach is the work by Rajh et al. that models wood-fired grate boiler [26]. An in-house code was used to determine the surface reaction flux fed to the CFD model's boundary. Recycled Flu Gas (RFG) jets, secondary, and tertiary air nozzles were used as input boundaries for other gas phase reactants. RANS CFD was combined with EDC for combustion modeled with a simple kinetic mechanism, like in the other examples.

The examples that followed this approach had acceptable fluid flow analysis fidelity, less detailed gas reaction model accuracy, and lower computational power requirements, all at the cost of excluding surface reactions from the model. This would imply that that surface reaction knowledge would have already been established from experimental research, which limits the investigation of wood species and geometry on the stove output. Moreover, combining the rather simplified chemical kinetic mechanisms with RANS time-averaged turbulence models results in lower fidelity combustion modeling. While this approach is reasonable for general performance investigations, basic emission analysis, and geometric design optimization, it falls short on predicting the onset of pollutants formation.

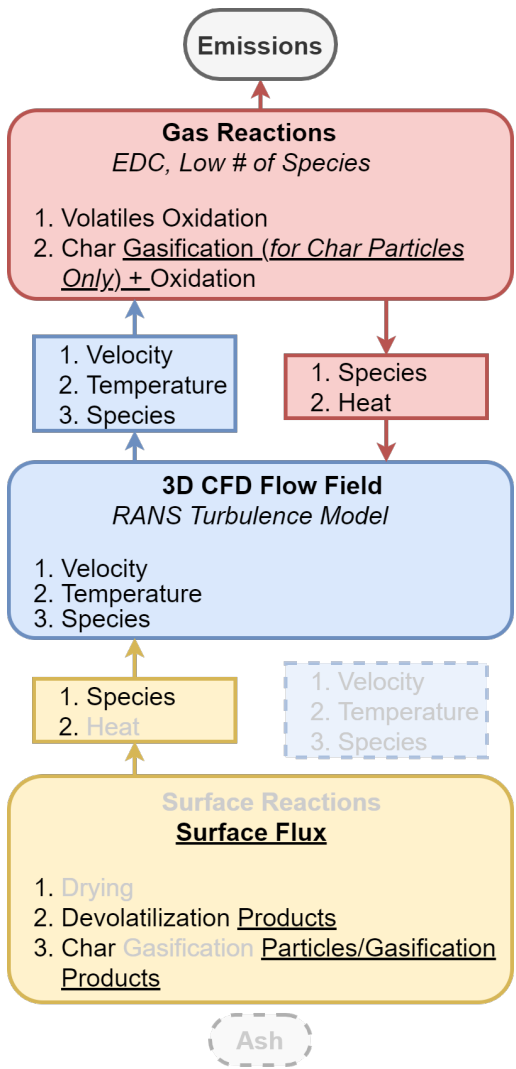

**Figure 5**: Model 1 diagram for Surface Flux → 3D RANS CFD Flow ↔ EDC Combustion Model [10, 24, 25, 26].

### 3.2. Surface Reactions → 0D Control Volume Flow ⇌ Arrhenius Combustion Model

In this approach, shown in the schematic of Figure 6, pyrolysis is modeled as a surface reaction, generating an output of drying and devolatilization products. Char particles are emitted to the fluid domain without a gasification surface reaction. For flow analysis, no CFD or turbulence modeling is done, but rather a Control Volume analysis transfers species and energy from the surface reaction model to the gas reaction model, and from the gas reaction model to the system boundaries.

Ritcher et al. used this approach to measure the combustion of a wood-fired hydronic heater. Using a pyrolysis to model surface reactions of wood allowed them to track the reactant species over the solution time domain. They then used CV analysis to transfer species and reactants to an Arrhenius combustion model. The results from combustion were transferred to system boundaries via the CV surrounding it [27].

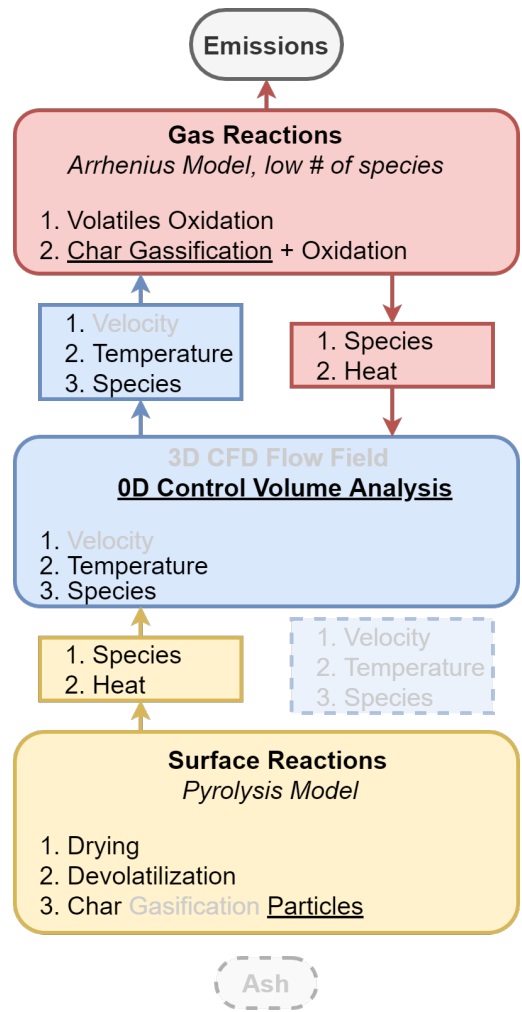

**Figure 6**: Model 2 diagram for Surface Reactions → 0D Control Volume Flow ⇌ Arrhenius Combustion Model [27].

The biggest advantage of this approach is the use of a surface reaction model, which allows for changes in chemical species and pollutants to be tracked and analyzed over the solution time. However, without an actual CFD flow simulation model, the gas combustion model will not account for effects of turbulence on chemical kinetic reaction rates. Additionally, the domain will have no spatial data resolution such as for temperature or species distributions.

### 3.3. Surface Reactions ↔ 3D RANS CFD Flow ↔ EDM Combustion

The approach shown in the schematic of Figure 7 is quite similar to the previous approach, with one main improvement, the use of a CFD model. Scharler et al. modeled surface reactions for a wood log, generating a time dependent flux of drying, devolatilization and gasification products [18]. CFD with RANS turbulence modeling (RNG k-$\varepsilon$) was used for flow simulation. The flow model was coupled with EDM for combustion modeling.

This approach is one of the most robust ones to model cord wood combustion with acceptable accuracy, as it features a reasonable combination of flow, combustion, and transient surface reaction modeling. To increase the model's fidelity, a more detailed turbulence model and a higher accuracy combustion model with a larger number of species could be added.

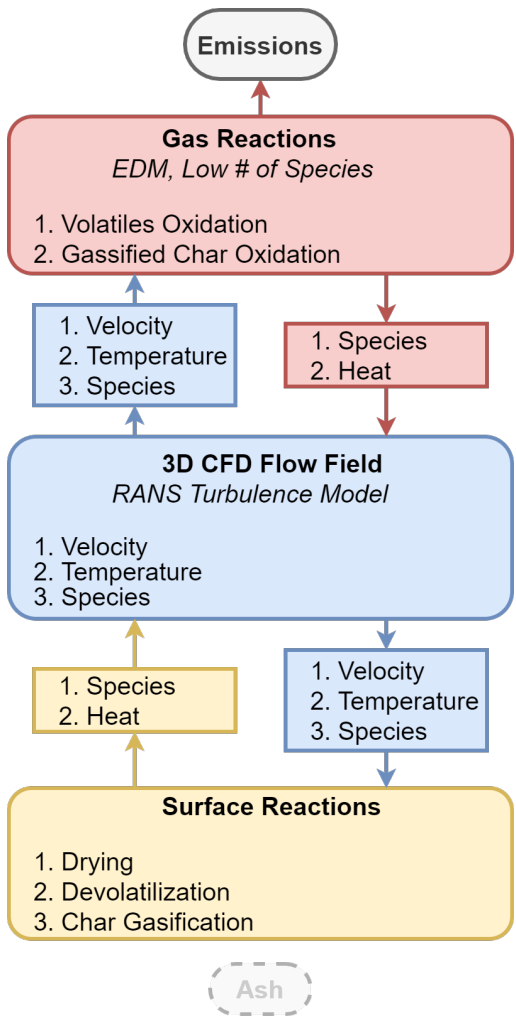

**Figure 7:** Model 3 diagram for Surface Reactions ↔ 3D RANS CFD Flow ↔ EDM Combustion [18].

### 3.4. DEM + Surface Reactions ↔ 3D RANS CFD Flow ↔ EDC Combustion

The approach shown in the schematic of Figure 8 is quite similar to the previous one, described in Section 3.3. However, for small wood pellets, the domain boundaries can never be dynamic and at the same time fine enough to capture the movement and geometries of wood pellets of different sizes and shapes. Therefore, a Discrete Element Method (DEM) is used to capture wood pellets, represented as point masses with spherical volumes to reduce dimensionality and decrease computational requirements. The DEM model is fully coupled to a CFD model with a RANS turbulence model, which – in turn, is fully coupled to an EDC combustion model.

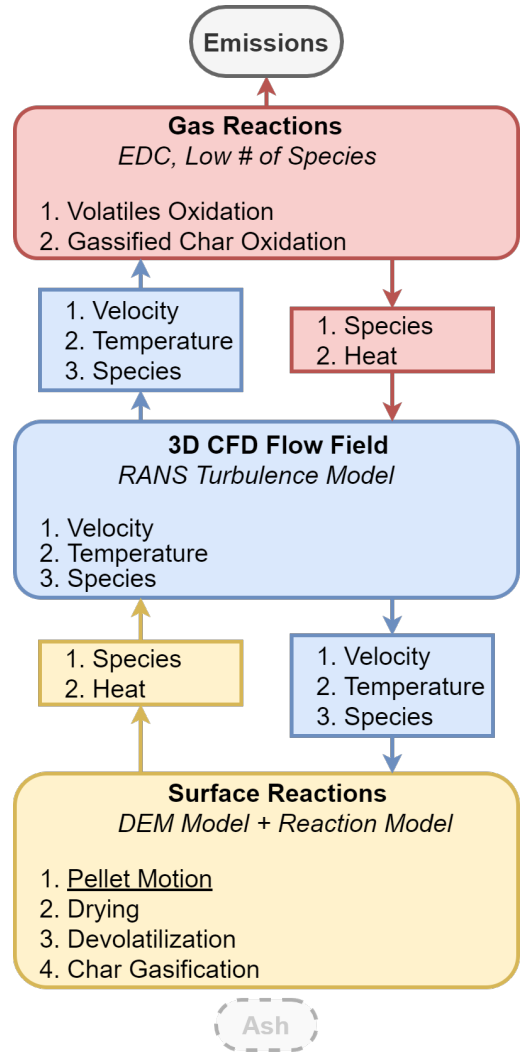

**Figure 8:** Model 4 diagram for DEM + Surface Reactions ↔ 3D RANS CFD Flow ↔ EDC Combustion [23].

Wiese et al. used this approach to model pellet wood stoves with a high degree of accuracy [23]. A DEM model representing simple spherical pellets was combined with various RANS turbulence models in the CFD code and an EDC combustion model, that featured a chemical kinetic mechanism with four (4) reactions and five (5) species.

Like the previous approach in Section 3.3, this model combines CFD, combustion and surface reaction modeling. However, RANS cannot fully capture flow structures and hence EDC can not fully model the effects of turbulence on chemical reaction rates. Moreover, the chemical kinetic mechanism used in this model is not detailed enough for high-fidelity applications. Moreover the DEM model uses spherical elements to represent pellets, which is computationally efficient, but it neglects the effect of the real non-uniform cylindrical shape of the pellets on surface reaction and pellet motion.

### 3.5. Surface Flux ⇌ 0D/1D Empirical CFB Flow ⇌ Arrhenius Combustion

The approach shown in the schematic of Figure 9 is used to model wood chips CFB's, commonly found in larger industrial settings. From a modeling perspective, it is worth exploring alternative approaches that can potentially inform wood stove and other residential wood combustion modeling techniques. Like the previous approach, surface reactions are replaced with species flux. However, the wood chip particles move along the flow field in a CFB, and the surface flux emitted from them varies with the local wood chip concentration. No turbulence modeling of the flow field is used. Rather, a direct analytical solution for the flow domain is coupled with two-phase flow equations to model the wood particles. The flow field is coupled with an Arrhenius combustion model, incorporating a more detailed chemical kinetic mechanism with a larger number of species. While CFB work is not directly tied to wood stove analysis, the more detailed chemical kinetic mechanisms used in CFB literature provide a good resource for wood stove modeling research.

Examples of this approach include Gungor's work, which modeled a 2D CFB using N-S equations for a two-phase flow [20]. Surface reactions were replaced by a variable surface flux, with intensity tied to wood chip concentration. Gas reactions were modeled with a more detailed Arrhenius chemical reaction model featuring fifteen (15) species and twenty four (24) reactions.

Adanez et al. used a similar approach to model a wood chip Fired CFB, with the exception of modeling the two-phase flow as a 0D CV instead of a 2D domain [28]. Their surface flux model included variable wood chip mass to account for reduction in particle mass and surface flux due to drying, devolatilization and gasification. The gas reactions used an eight (8) species and 5 reactions Arrhenius chemical reaction model.

This methodology has the benefits of much faster computational times, and the capability of generating solutions over a much larger time domain. Moreover, the CFB model's inclusion of the local concentration of wood chips greatly increases the surface flux accuracy. However, the surface flux is not as accurate as could be predicted from models incorporating surface reactions. The lack of the use of a CFD model further compromises the prediction of gas reactions, as the effects of turbulence on chemistry are not captured.

## 4. Current Gaps in Wood Modeling Research

Current gaps in wood modeling research are due to two main issues. The first is that the physical phenomena to be modeled are complex and each component of a wood combustion model has a different time scale. For example, a piece of cord wood takes a large amount of time (minutes) to fully go through pyrolysis, leaving only ash after burning. On the other hand, turbulence in the flow varies at a time scale that is much smaller (fractions of a second). Finally, volatile gases go through a complete combustion reaction on a time scale that is orders of magnitude smaller than turbulence. The second issue is the need to compromise between model fidelity of physical phenomena and computational power requirements.

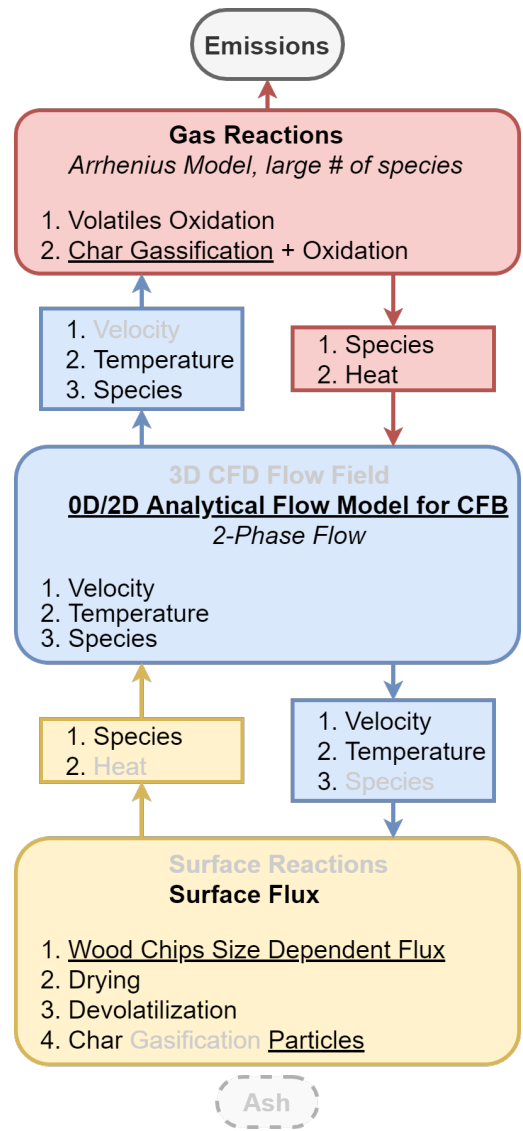

**Figure 9:** Model 5 diagram for Surface Flux ⇌ 0D/1D Empirical CFB Flow ⇌ Arrhenius Combustion [20, 28].

Empirical reaction models are less computationally demanding since they mostly depend on existing experimental data. However, they do not provide any new information about the chemical reaction details. Rather, such models commonly serve as tools to generate boundary conditions for the CFD domain. This approach could be sufficient to investigate different stove designs and determine spatially-averaged temperature distributions; however, it is neither detailed enough to explore spatially-resolved parameters (e.g., flow-field velocity, temperature distribution, and chemical species concentrations), nor comprehensive enough to model detailed chemical reactions (e.g., the effects of wood composition on combustion or pollutant formation).

Although 0D and CV flow models are usually coupled with higher fidelity chemical kinetic mechanisms, the CV approach lacks spatial resolution of flow effects on combustion. Since turbulence effects are only computed empirically, this limits the application of the model to predicting only bulk domain variables, such as those used for system sizing (e.g., thermal output).

With the use of RANS equations for turbulence modeling. The turbulence in the flow field is not modeled at detailed time scales, but rather averaged and dampened over the solution time. Since combustion occurs between reactants at a time scale that is order of magnitudes smaller than flow time scales, and due to the dependence of detailed combustion models on temperature, species, and mixing and turbulence, the RANS approach to resolving turbulence is not enough to provide flow field data accurate enough for combustion modeling.

CFD challenges rise with dynamic boundaries (e.g., deformation of a burning wood log) since they require a more complex meshing process that conforms with any deforming or moving system boundaries. Models coupled with DEM provide a higher fidelity simulation of surface reactions and fuel particle movement of wood chips and pellets in the domain. While such models can accurately relay their data to the boundary conditions of the CFD model, the latter still lacks in combustion resolution due to the dependence on RANS.

The current EDC chemical kinetics-based reaction models that are coupled with 3D RANS CFD models tend to use relatively simplified reaction models, with simplistic boundary conditions, and reduced chemical kinetic mechanisms (average of eight (8) species). Moreover, they are coupled with RANS turbulence modeling which cannot unlock the potential of EDC combustion modeling. Although current combustion models can predict species concentration at system boundaries (e.g., exhaust stack), they typically cannot resolve such details in the 3D domain.

The majority of the issues identified above can be mitigated by using more sophisticated models that require additional computational power and detailed experimental data for model development and validation (e.g., better detailed chemical kinetic mechanisms, higher fidelity dynamic boundary conditions/DEM models).

## 5. Proposed Approach for Modeling Wood Combustion

The state-of-the-art in modeling and experimental testing of wood stove combustion can typically provide only time-averaged flow field information and chemical species data at the system boundaries (e.g., air intake ports and/or exhaust stack). Moreover, understanding where emissions pollutants (e.g., PM, CO, UHC, $CH_4$, $NO_x$) are formed spatially inside the stove and under what specific localized conditions remains out of reach with existing computational modeling tools. In-situ experimental measurements under such harsh, high-temperature, and particle-laden environments are also extremely challenging. Accurately modeling wood combustion in wood stoves requires a multi-stage process.

The initial step is to develop a thorough understanding of the physical phenomena supplemented by detailed experimental data. Accurate reaction modeling starts by developing and using detailed surface and gas reaction chemical kinetic models for drying, devolatilization, gasification and oxidation. The chemical kinetic models should be thoroughly tested using 0D simulation tools (e.g., Cantera, CHEMKIN). Then the simulations should be validated against detailed experimental data. Published work with more detailed chemical kinetic mechanisms will further enable the development of this stage. Additionally, the boundaries for the CFD model should be modeled accurately to capture changes in wood geometry, mass, and composition during combustion, effectively creating a robust surface reaction model. Moreover, tools like NIST's Fire Dynamic Simulator (FDS), which are used to model Fire and smoke propagation in detail [29, 30], can aid in investigating pollutants and reacting species in the flow.

Finally, the CFD model should be capable of capturing the time-dependent nature of wood burning. Both surface and gas phase reactions should be modeled using multi-step kinetic mechanisms with the use of detailed chemistry combustion models (e.g., SAGE). Moreover, turbulence should be modeled to a higher degree of accuracy. In particular, the use of LES turbulence modeling can provide a more detailed flow field, capturing larger eddies and flow structures accurately. Such turbulence modeling technique provides a CFD model that scales in accuracy with finer computational meshes. Modern commercial CFD codes (e.g., CONVERGE CFD) can now use automatic adaptive cut-cell meshing to manage dynamic system boundaries, similar to Sofianopoulos' et al. work that investigated thermal stratification in homogeneous charge compression ignition engines and Guleria's et al. work that investigated stratified charge compression ignition work using LES modeling [31, 32]. It is noteworthy to mention that one of the biggest challenges in using LES to model wood stoves comes in finding a computationally efficient method to address the large solution time of wood stoves that span minutes or hours compared to that of traditional combustion applications (i.e. internal combustion engines) which span seconds.

Figure 10 shows our proposed approach for modeling wood stove combustion with higher fidelity. This can be achieved by modeling different regions of the combustion domain for a few seconds in a multi-zone manner. This data could be distributed, giving accurate results over the larger time domain, while capturing the smaller time scale data and fluctuations. The combination of detailed dynamic boundary conditions or DEM, detailed chemical kinetic mechanisms, robust combustion models, and LES turbulence modeling can provide a much more accurate analysis of wood combustion, with detailed spatially-resolved flow field and species data.

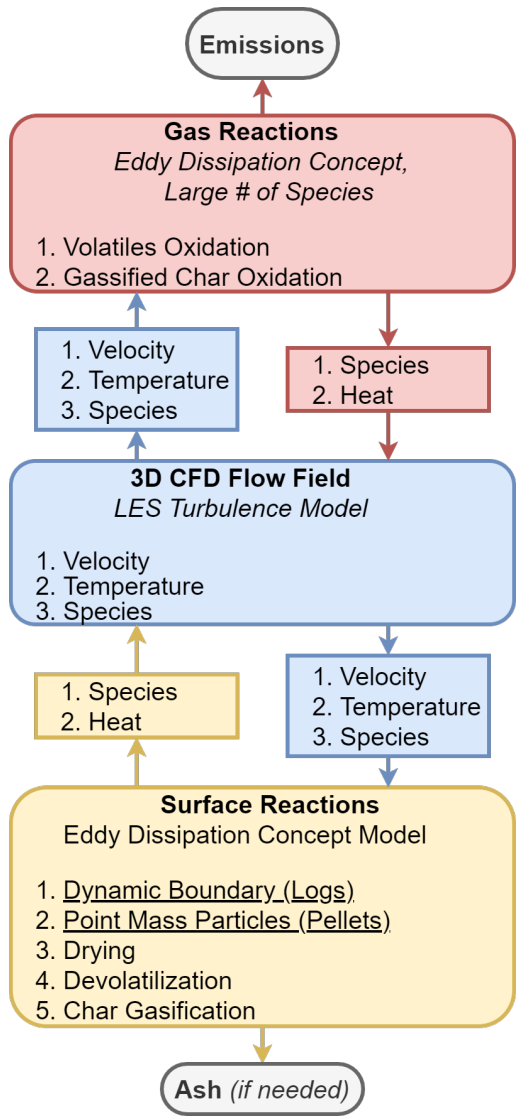

**Figure 10:** Proposed model for wood combustion with detailed combustion and LES CFD modeling.

The salient features of all the different models are summarized and compared in Table 1. Depending on desired application needs and computational resources available, the use of certain models may be more appropriate than others. The 0D and 2D models have low computational requirements and can reasonably determine fluxes at the stove boundaries (e.g., heat transfer in/out, emissions), but lack the ability to provide data resolution within the stove computational domain (e.g., species concentration, temperature distribution, turbulence); this makes them good candidates for use in system sizing problems. On the other hand, the 3D RANS CFD models offer a balance between computational power requirements, ability to calculate field variables inside the stove domain, and fluxes crossing the system boundaries. While there use for general design studies requiring multiple iterations is appropriate, the computational domain field variables are time-averaged and hence cannot provide detailed instantaneous data. Finally, the proposed 3D LES CFD model can be very expensive computationally, as it requires multiple computational cycles to calculate system boundary fluxes at reasonable accuracy; this renders it not that viable for system sizing problems. However, field variables can be resolved in both time and space at relatively high accuracy, allowing for a detailed understanding of how all the elements inside the stove contribute to pollutant formation and system efficiency. This approach can provide a better understanding in-stove phenomena and help in detailed design optimization.

## 6. Other Model Considerations to Reflect Realistic Usage

Real wood stove operation and wood combustion can vary widely from model predictions that ultimately reflect idealized wood stove and wood combustion operation. Specifically some of these factors that are not captured by most models and can vary in real world operation of wood combustion devices include:

- Wood species (hardwoods vs. softwoods)
- Geometry of wood (crib wood vs. cord wood vs. log wood)
- Presence of exterior bark
- Overall wood moisture content
- Internal moisture content variations
- Residual inorganic (e.g., ash) or organic (e.g., creosoot) matter in combustion chamber and/or exhaust stack

Such factors must be considered and related sub-models, where possible, should be developed as we keep improving overall wood combustion model fidelity. Lastly, these factors can also play a critical role in emissions generated, specifically nanoparticle emissions [33], which are much more difficult to remove using traditional emission control devices from wood combustion exhaust streams.

## 7. Conclusions

A pressing need exists – now more than ever - for a state-of-the-art, high-fidelity, multi-dimensional numerical model that can accurately represent the physical processes that occur within a wood stove (surface reactions, fluid flow, and gas reactions). The model should have sufficient resolution to provide detailed information inside the computational domain, not just at the boundaries. Once such a model is validated, it will be able to capture details within the stove domain, such as the onset of NOx formation or regions of high soot and CO concentrations, that are not measurable via traditional experimental techniques, which are typically applied at the boundaries.

**Table 1**
Comparison of different wood combustion models reviewed and proposed 3D LES higher-fidelity model.

| | Model 1<br>Surface Flux →<br>3D RANS CFD Flow ↔<br>EDC Combustion Model | | Model 2<br>Surface Reactions →<br>0D Control Volume Flow ⇌<br>Arrhenius Combustion Model | | Model 3<br>Surface Reactions ↔<br>3D RANS CFD Flow ↔<br>EDM Combustion | | Model 4<br>DEM + Surface Reactions ↔<br>3D RANS CFD Flow ↔<br>EDC Combustion | | Model 5<br>Surface Flux ⇌<br>0D/1D Empirical CFB Flow ⇌<br>Arrhenius Combustion | | Proposed<br>EDC Surface Reactions ↔<br>3D LES CFD Flow ↔<br>EDC Gas-Phase Combustion | |
|---|---|---|---|---|---|---|---|---|---|---|---|---|
| Surface Model | Boundary Flux (Arrhenius) | + | No | - | Deforming Boundary | ++ | DEM Moving / Deforming Pellet | ++ | No | - | Deforming Boundary / DEM Moving Pellets | ++ |
| Reaction Model | EDC | ++ | Arrhenius | - | EDM | + | EDC | ++ | Arrhenius | - | EDC | ++ |
| Number of Species | Low | - | Low | - | Low | - | Low | - | Moderate | + | Large | ++ |
| Drying | No | - | Yes | + | Yes | + | Yes | + | Yes | + | Yes | + |
| Devolatilization | Volatiles Flux | - | Volatiles Production | + | Volatiles Production | + | Volatiles Production | + | Volatiles Production | + | Volatiles Production | + |
| Char Gasification | Char Particle Flux | - | Char Particle Flux | - | Char Gasification | + | Char Gasification | + | Char Particle Flux | - | Char Gasification | + |
| Flow Field | 3D CFD | ++ | 0D CV | - | 3D CFD | ++ | 3D CFD | ++ | 0D / 2D Analytical 2-Phase Model | + | 3D CFD | ++ |
| Turbulence Model | RANS | + | N/A | - | RANS | + | RANS | + | N/A | - | LES | ++ |
| Surface to Flow Coupling | 1-way | - | 1-way | - | 2-way | ++ | 2-way | ++ | Partial 2-way | + | 2-way | ++ |
| Gas Reaction to Flow Coupling | 2-way | ++ | Partial 2-way | + | 2-way | ++ | 2-way | ++ | Partial 2-way | + | 2-way | ++ |
| Ability to Resolve Data at Boundaries (Stove Sizing, Emissions Out, Efficiency) | ++ | | ++ | | ++ | | ++ | | ++ | | ++ | |
| Ability to Resolve Inside-Stove Phenomena (Emission Onsets, Surface Deformation, Detailed Species Concentration) | + | | - | | + | | + | | - | | ++ | |
| Computational Requirements (Efficiency) | + | | +++ | | + | | + | | ++ | | - | |

The other important aspect of developing such a high-fidelity advanced model is that it should not be restricted to wood combustion only. This way, the use of this methodology could be extended to other important applications with detailed surface and gas reactions in a complex flow field, such as catalytic reactors or other types of biomass and solid fuel combustion.

In parallel to the development of advanced numerical modeling approaches, experimental measurement techniques are required to develop insights into the phenomena and validate the models. This could include the use of high-speed cinematography to capture flow motion and combustion chemiluminescence through optical chambers, similar to the previous works of Assanis et al. [34, 35]; or the use of Particle Image Velocimetry, which can provide much detailed insight to in-oven flow and reactions [36].

The combined use of advanced, high-fidelity models and experimental techniques can enable Computer Aided Engineering of wood combustion, thus creating a paradigm shift in the development time required to reduce emissions and provide cleaner and more efficient burning stoves sooner. While computational power requirements will increase, High Performance Computing facilities are increasingly becoming available to researchers and thus now making this paradigm shift possible.

# References

[1] US Energy Information Administration. Increase in wood as main source of household heating most notable in the northeast. https://www.eia.gov/todayinenergy/detail.php?id=15431, 2014. Accessed: 2021-02-13.

[2] US Energy Information Administration. Residential energy consumption survey (recs). https://www.eia.gov/consumption/residential/data/2015/, 2015. Accessed: 2021-02-13.

[3] US Census Bureau. Household income for states: 2008 and 2009. https://www2.census.gov/library/publications/2010/acs/acsbr09-02.pdf, 2010. Accessed: 2021-02-13.

[4] L. Weiss et al. New York State Wood Heat Report: An Energy, Environmental, and Market Assessment. *New York State Energy Research and Development Authority (NYSERDA) Report 15-26*, 2016.

[5] J.B. Heywood. *Internal Combustion Engine Fundamentals*. McGraw-Hill Education, 1988.

[6] E. Mutlu, S. H. Warren, S. M. Ebersviller, I. M. Kooter, J. E. Schmid, J. A. Dye, W. P. Linak, M. I. Gilmour, J. J. Jetter, M. Higuchi, and D. M. DeMarini. Mutagenicity and pollutant emission factors of solid-fuel cookstoves: Comparison with other combustion sources. *Environmental Health Perspectives*, 124(7):974–982, jul 2016. doi: 10.1289/ehp.1509852.

[7] L. Soto-Garcia, W. J. Ashley, S. Bregg, D. Walier, R. LeBouf, P. K. Hopke, and A. Rossner. VOCs emissions from multiple wood pellet types and concentrations in indoor air. *Energy & Fuels*, 29(10):6485–6493, sep 2015. doi: 10.1021/acs.energyfuels.5b01398.

[8] US Environmental Protection Agency. Burn wise. https://www.epa.gov/burnwise, 2021. Accessed: 2021-02-13.

[9] T. Palander, H. Haavikko, and K. Kärhä. Towards sustainable wood procurement in forest industry – the energy efficiency of larger and heavier vehicles in finland. *Renewable and Sustainable Energy Reviews*, 96:100–118, nov 2018. doi: 10.1016/j.rser.2018.07.043.

[10] P. Motyl, M. Wikło, J. Bukalska, B. Piechnik, and R. Kalbarczyk. A new design for wood stoves based on numerical analysis and experimental research. *Energies*, 13(5):1028, feb 2020. doi: 10.3390/en13051028.

[11] T. Nussbaumer. Combustion and co-combustion of biomass: fundamentals, technologies, and primary measures for emission reduction†. *Energy & Fuels*, 17(6):1510–1521, nov 2003. doi: 10.1021/ef030031q.

[12] L. Marabini, S. Ozgen, S. Turacchi, S. Aminti, F. Arnaboldi, G. Lonati, P. Fermo, L. Corbella, G. Valli, V. Bernardoni, M. Dell'Acqua, R. Vecchi, S. Becagli, D. Caruso, G. L. Corrado, and M. Marinovich. Ultrafine particles (UFPs) from domestic wood stoves: genotoxicity in human lung carcinoma a549 cells. *Mutation Research/Genetic Toxicology and Environmental Mutagenesis*, 820:39–46, aug 2017. doi: 10.1016/j.mrgentox.2017.06.001.

[13] R. Rao and M. Esposito. Nonequilibrium thermodynamics of chemical reaction networks: Wisdom from stochastic thermodynamics. *Physical Review X*, 6(4), dec 2016. doi: 10.1103/physrevx.6.041064.

[14] J. Krüger, N. Erl, L. Haugen, and T. Løvås. Correlation effects between turbulence and the conversion rate of pulverized char particles. *Combustion and Flame*, 185:160–172, nov 2017. doi: 10.1016/j.combustflame.2017.07.008.

[15] M. Muto, H. Watanabe, R. Kurose, S. Komori, S. Balusamy, and S. Hochgreb. Large-eddy simulation of pulverized coal jet flame – effect of oxygen concentration on NOx formation. *Fuel*, 142:152–163, feb 2015. doi: 10.1016/j.fuel.2014.10.069.

[16] S. Som, P.K. Senecal, and E. Pomraning. Comparison of rans and les turbulence models against constant volume diesel experiments. In *ILASS Americas, 24th annual conference on liquid atomization and spray systems, San Antonio, TX*, 2012.

[17] E.R. Vasquez and T. Eldredge. 18 - process modeling for hydrocarbon fuel conversion. In M. Rashid Khan, editor, *Advances in Clean Hydrocarbon Fuel Processing*, Woodhead Publishing Series in Energy, pages 509–545. Woodhead Publishing, 2011. ISBN 978-1-84569-727-3. doi: https://doi.org/10.1533/9780857093783.5.509. URL https://www.sciencedirect.com/science/article/pii/B9781845697273500184.

[18] R. Scharler, T. Gruber, A. Ehrenhöfer, J. Kelz, R. M. Bardar, T. Bauer, C. Hochenauer, and A. Anca-Couce. Transient CFD simulation of wood log combustion in stoves. *Renewable Energy*, 145:651–662, jan 2020. doi: 10.1016/j.renene.2019.06.053.

[19] D.D. Toporov. Chapter 4 - mathematical modelling and model validations. In D.D. Toporov, editor, *Combustion of Pulverised Coal in a Mixture of Oxygen and Recycled Flue Gas*, pages 51–97. Elsevier, Boston, 2014. ISBN 978-0-08-099998-2. doi: https://doi.org/10.1016/B978-0-08-099998-2.00004-7. URL https://www.sciencedirect.com/science/article/pii/B9780080999982000047.

[20] A. Gungor. Two-dimensional biomass combustion modeling of CFB. *Fuel*, 87(8-9):1453–1468, jul 2008. doi: 10.1016/j.fuel.2007.08.013.

[21] A. Anca-Couce, G. Caposciutti, T. Gruber, J. Kelz, T. Bauer, C. Hochenauer, and R. Scharler. Single large wood log conversion in a stove: Experiments and modelling. *Renewable Energy*, 143:890–897, dec 2019. doi: 10.1016/j.renene.2019.05.065.

[22] A. Trubetskaya, G. Beckmann, J. Wadenbäck, J. K. Holm, S. P. Velaga, and R. Weber. One way of representing the size and shape of biomass particles in combustion modeling. *Fuel*, 206:675–683, oct 2017. doi: 10.1016/j.fuel.2017.06.052.

[23] J. Wiese, F. Wissing, D. Höhner, S. Wirtz, V. Scherer, U. Ley, and H. M. Behr. DEM/CFD modeling of the fuel conversion in a pellet stove. *Fuel Processing Technology*, 152:223–239, nov 2016. doi: 10.1016/j.fuproc.2016.06.005.

[24] S. Kalla, H. Marcoux, and A. Champlain. Cfd approach for modeling high and low combustion in a natural draft residential wood log stove. *International Journal of Heat and Technology*, 33(1):33–38, mar 2015. doi: 10.18280/ijht.330105.

[25] A. Nesiadis, N. Nikolopoulos, N. Margaritis, P. Grammelis, and E. Kakaras. Optimization of a log wood boiler through CFD simulation methods. *Fuel Processing Technology*, 137:75–92, sep 2015. doi: 10.1016/j.fuproc.2015.04.010.

[26] B. Rajh, C. Yin, N. Samec, M. Hriberšek, F. Kokalj, and M. Zadravec. Advanced CFD modelling of air and recycled flue gas staging in a waste wood-fired grate boiler for higher combustion efficiency and greater environmental benefits. *Journal of Environmental Management*, 218:200–208, jul 2018. doi: 10.1016/j.jenvman.2018.04.030.

[27] J.P. Richter, J.M. Weisberger, B.T. Bojko, J.C. Mollendorf, and P.E. DesJardin. Numerical modeling of homogeneous gas and heterogeneous char combustion for a wood-fired hydronic heater. *Renewable Energy*, 131:890–899, feb 2019. doi: 10.1016/j.renene.2018.07.087.

[28] J. Adánez, P. Gayán, L. F. de Diego, F. García-Labiano, and A. Abad. Combustion of wood chips in a CFBC. modeling and validation. *Industrial & Engineering Chemistry Research*, 42(5):987–999, mar 2003. doi: 10.1021/ie020605z.

[29] S. Zhang, X. Ni, M. Zhao, J. Feng, and R. Zhang. Numerical simulation of wood crib fire behavior in a confined space using cone calorimeter data. *Journal of Thermal Analysis and Calorimetry*, 119(3):2291–2303, dec 2014. doi: 10.1007/s10973-014-4291-4.

[30] J. Wahlqvist and P. van Hees. Validation of FDS for large-scale well-confined mechanically ventilated fire scenarios with emphasis on predicting ventilation system behavior. *Fire Safety Journal*, 62:102–114, nov 2013. doi: 10.1016/j.firesaf.2013.07.007.

[31] A. Sofianopoulos, M. Rahimi, B. Lawler, and S. Mamalis. Investigation of thermal stratification in premixed homogeneous charge compression ignition engines: A large eddy simulation study. *International Journal of Engine Research*, 20:146808741879552, 09 2018. doi: 10.1177/1468087418795525.

[32] G. Guleria, D. Lopez-Pintor, J. E. Dec, and D. Assanis. A comparative study of gasoline skeletal mechanisms under partial fuel stratification conditions using large eddy simulations. *International Journal of Engine Research*, 2021.

[33] R. Trojanowski and V. Fthenakis. Nanoparticle emissions

from residential wood combustion: A critical literature review, characterization, and recommendations. *Renewable and Sustainable Energy Reviews*, 103:515–528, 2019. ISSN 1364-0321. doi: https://doi.org/10.1016/j.rser.2019.01.007. URL https://www.sciencedirect.com/science/article/pii/S1364032119300012.

[34] D. Assanis, N. Engineer, P. Neuman, and M.S. Wooldridge. Computational development of a dual pre-chamber engine concept for lean burn combustion. Technical report, SAE Technical Paper 2016-01-2242, 2016.

[35] D. Assanis, S.W. Wagnon, and M.S. Wooldridge. An experimental study of flame and autoignition interactions of iso-octane and air mixtures. *Combustion and Flame*, 162(4), 4 2015. doi: 10.1016/j.combustflame.2014.10.012.

[36] F. Chen and H. Liu. Particle image velocimetry for combustion measurements: Applications and developments. *Chinese Journal of Aeronautics*, 31(7):1407–1427, jul 2018. doi: 10.1016/j.cja.2018.05.010.